\documentclass[journal]{IEEEtran}

\usepackage{url}

\usepackage{xcolor,cite,enumerate}
\usepackage{amssymb}
\usepackage{amsmath}
\usepackage{graphicx}
\usepackage{subfigure,balance}
\usepackage{booktabs}
\usepackage{balance}
\usepackage{multirow}
\usepackage[inline]{enumitem}

\begin{document}

\title{Crossterm-Free Time-Frequency Representation Exploiting Deep Convolutional Neural Network}

\author{Shuimei Zhang, \IEEEmembership{Student Member, IEEE}, and Yimin D.\ Zhang, \IEEEmembership{Fellow, IEEE}
\thanks{The authors are with the Department of Electrical and Computer Engineering, Temple University, Philadelphia, PA 19122.}  }

\markboth{Crossterm-Free Time-Frequency Representation}
{Shell \MakeLowercase{\textit{et al.}}: Bare Demo of IEEEtran.cls for IEEE Journals}
\maketitle

\begin{abstract}
Bilinear time-frequency representations (TFRs) provide high-resolution time-varying frequency characteristics of nonstationary signals. However, they suffer from crossterms due to the bilinear nature. Existing crossterm-reduced TFRs focus on optimized kernel design which amounts to low-pass weighting or masking in the ambiguity function domain. Optimization of fixed and adaptive kernels are difficult, particularly for complicated signals whose autoterms and crossterms overlap in the ambiguity function. In this letter, we develop a new method to offer  high-resolution TFRs of nonstationary signals with crossterms effectively suppressed. The proposed method exploits a deep convolutional neural network which is trained to construct crossterm-free TFRs. The effectiveness of the proposed method is verified by simulation results which clearly show desirable autoterm preservation and crossterm mitigation capabilities. The proposed technique significantly outperforms state-of-the-art time-frequency analysis algorithms based on adaptive kernels and compressive sensing techniques.
\end{abstract}


\IEEEpeerreviewmaketitle

\section{Introduction}

\IEEEPARstart{N}{onstationary} signals are naturally observed in various real-world applications, such as radar, sonar, satellite navigation, and biomedical applications \cite{application_book, chen_book, asilomar07, stankovic_book, spm_tf, boashash_book, spm17, eeg, Shui_radar}. One important class of nonstationary signals is characterized by time-varying instantaneous frequencies (IFs) with constant magnitude or time-varying magnitudes. These signals are respectively referred to as frequency-modulated (FM) signals. For such signals, joint time-frequency (TF) domain representations are most suited for their analyses and classification as they effectively provide time-varying spectra along the true signal IFs.

TF representations (TFRs) can be generally classified into linear and bilinear. For example, short-time Fourier transform is a commonly used linear TFR whose TF resolution is restricted to a fixed sliding time window. Compared to the linear counterparts, bilinear TFRs generally provide higher TF concentration. However, due to the nonlinearity of the bilinear TFRs, crossterms unavoidably appear midway between signal autoterm components in the case of nonlinear or multi-component signals. Such crossterms prohibit accurate analysis and interpretation of the signal IF signatures \cite{ boashash_book}.

The Wigner-Ville distribution (WVD) is often referred to as the prototype bilinear TFR with a high impact of crossterms. Various TF kernels have been developed to suppress crossterms while preserving autoterms \cite{stankovic_book, boashash_book, Jones_AOK,adtfd}. Essentially, a TF kernel acts as a two-dimensional (2-D) low-pass filter or mask multiplied in the ambiguity function (AF) domain, expressed with respect to time lag and frequency shift, because typically autoterms have high values around the origin of the AF domain whereas crossterms tend to be away from the origin. Effectively, a kernel becomes a 2-D convolution in the TF domain. There is a trade-off between autoterm preservation and crossterm mitigation. Designing a TF kernel function with satisfactory autoterm preservation and crossterm suppression has been a challenging task for many decades and motivated the development of a high number of TF kernels.
Existing kernel designs generally assume that crossterms are well separated from signal autoterms in the AF domain. However, this assumption is not always valid especially when the signal involves complex structure, such as signals with highly nonlinear and intersecting TF signatures.

Because FM signals are sparsely presented in the TF domain, obtaining high-resolution TFRs with reduced crossterms can also be formulated as a sparse reconstruction problem. The incorporation of TF sparsity has been implemented in two ways. The first class of approaches utilizes the 2-D Fourier transform relationship between the AF and TFR \cite{Flandrin_time, Flandrin2015, Deprem_cross, Jiang_2020}. In these methods, a proper mask around the AF origin is selected to mitigate the effects of the crossterms. The desired TFR is then found by minimizing the $\ell_1$-norm of the vectorized TFR entries. On the other hand, the second class of approaches is based on the one-dimension (1-D) Fourier transform that relates the instantaneous auto-correlation function (IAF) and the TFR, and sparsity-based TFRs are obtained by performing compressive sensing on the prototype or kenneled IAF for each time instant \cite{Yimin_reduced, amin15, Shui_AT}. The complexity of the second class  of approaches is much lower than the first one because the dictionary matrix constructs 1-D Fourier transform rather than 2-D Fourier transform as in the first class. In addition, the second class provides better consistency of the TFR as it exploits local sparsity for each time instant.

Recently, deep learning \cite{cnn_review} has achieved a great success in many applications, such as image recognition \cite{image_recognition1},
speech recognition \cite{speech}, electroencephalogram (EEG) interpretation \cite{eeg_deep_learning}, crack detection \cite{crack}, human motion recognition \cite{wang}, and spectral recovery \cite{SAR}. Jiang \textit{et al} \cite{Jiang_2020} recently developed a U-Net aided iterative shrinkage-thresholding algorithm to learn the structural sparsity in the TF domain. However, its TFR reconstruction performance is still restricted by the selection of the AF samples.

\begin{figure*}[t]
  \centering
    \includegraphics[width=1\textwidth]{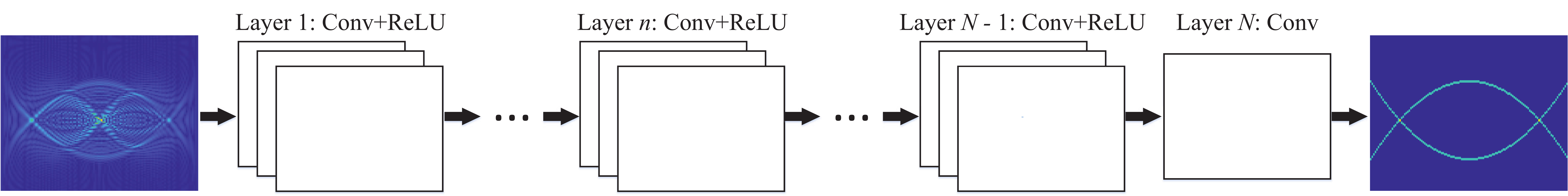}
  \caption{Proposed DCNN architecture to achieve the high-resolution crossterm-free TFR.}
  \label{CNN_structure2}
  \vspace{-1em}
\end{figure*}

In this letter, we develop a novel approach to obtain high-resolution crossterm-free TFRs. In particular, we exploit a deep convolutional neural network (DCNN) \cite{cnn_review} which is trained to extract the information of autoterms directly from the WVD. The ideal TF model is served as the training label. Unlike the existing kernel designs which design and optimize the kernel based on the signal characteristics observed in the AF domain, the proposed method offers an end-to-end learning fashion with loss function optimized at the network output. Therefore, it provides significant improvement in the the TFR reconstruction performance which does not suffer even autoterms and crossterms highly overlap in the AF.

\textit{Notations}$\colon$ Lower-case (upper-case) bold characters are used to denote vectors (matrices). ${\left(  \cdot  \right)^\mathrm{*}}$, ${\left(  \cdot  \right)^\mathrm{T}}$ and  ${\left(  \cdot  \right)^\mathrm{H}}$ denote complex conjugation, transpose and the Hermitian transpose, respectively. $\mathcal{F}_s(\cdot)$ represents the discrete Fourier transform (DFT) with respect to $x$. $\operatorname{Re}(\cdot)$ represents the real part of a complex value, and $\Vert \cdot \Vert_F$ denotes the Frobenius norm.

\vspace{-0.5em}
\section{Signal Model and Time-Frequency Representation}
Consider a discrete-time $P$-component FM signal
\begin{equation}
s(t)= \sum_{p=1}^{P} s_p(t) = \sum_{p=1}^{P}a_p e^{\jmath \phi_p(t)}, \ \ t= 1, \cdots, T,
\end{equation}	
where $a_p$ and $\phi_p(t)$ respectively denote the amplitude and the phase law of the $p$th component for $p=1,\cdots, P$. The IF of the $p$th signal component is given by $f_p(t) = {\rm d}\phi_p(t)/ (2\pi \ {\rm d}t)$.

The IAF of $s(t)$ is defined as
\begin{equation}
\begin{split} 	\label{IAF}
  	R_{ss}(t,\tau) &= s \left( t +\frac{\tau}{2} \right) s^* \left( t-\frac{\tau}{2} \right) \\
  	& = \sum_{p=1}^{P} s_{p} \left(t+\frac{\tau}{2}\right) s_{p}^{*}\left(t-\frac{\tau}{2}\right) \\
  	 & \quad  +\sum_{p=1 \atop }^{P} \sum_{q=1 \atop q \neq p}^{P} s_{p} \left(t+\frac{\tau}{2}\right)  s_{q}^{*}\left(t-\frac{\tau}{2}\right),
\end{split}
\end{equation}
where $\tau$ is the time lag.

The DFT of the IAF $R_{ss}(t,\tau)$ with respect to $\tau$ is the well-known WVD, i.e.,
\begin{equation} \label{wvd}
\begin{split}
	W_{ss}(t, f) & \!=\! \mathcal{F}_{\tau}[R_{ss}(t, \tau)] \\
	             &\!= \!\sum_{p=1}^{P}\! W_{s_p s_p}\!(t, f)\!+\!2 \sum_{p=1 \atop}^{P}\! \sum_{q=1  \atop q \neq p}^{P} \! \operatorname{Re}\!\left[W_{s_p s_q}(t, f)\right].
\end{split}
\end{equation}
The AF is obtained by applying 1-D DFT to the IAF $R_{ss}(t,\tau)$ with respect to $t$, expressed as
\begin{equation}
	A_{s s}(\theta, \tau)=\mathcal{F}_{t}[R_{s s}(t, \tau)], 
	\label{AF_def}
\end{equation}
where $\theta$ denotes the frequency shift or Doppler.
Clearly, the AF has a 2-D Fourier transform relationship with the WVD.

The last term in \eqref{wvd} represents the undesirable crossterm, which is a byproduct induced by the bilinear nature of the WVD and appear in the midway between autoterm components. TF kernls for crossterm mitigation are often implemented in the AF as a multiplicative filter emphasizing the region around the origin since, typically, autoterms are centered around the origin whereas the crossterms are dislocated from the origin. However, as the exact characteristics of autoterms and crossterms vary with each signal, various TF kernels have been designed. TF kernels can be classified into two general types, i.e., data-independent (fixed) and data-dependent (adaptive).

Adaptive kernels are designed to maximize certain performance measure under some constraints and generally provide better performance compared to fixed kernels. For example, the adaptive optimal kernel (AOK) \cite{Jones_AOK} optimizes the width of radial Gaussian function in different directions to maximize the weighted AF, whereas the adaptive directional TF distribution (ADTFD) \cite{adtfd} optimizes the direction of the smoothing kernel. These kernels do not have the exact knowledge of the autoterms and crossterms, and the optimization is limited to certain directions in the AF or TF domain. In addition, their performance deteriorates when the autoterms and crossterms are inseparable in the domain where the kernel is optimized.

\section{DCNN-based Crossterm-Free TFR}
In this section, we describe our proposed deep learning-based crossterm-free TFR by exploiting DCNN.
We view the crossterm-free TFR as a plain discriminative learning problem which, in essence, provides supervised TF kernel optimization capability to minimize the error at the network output. As such, the proposed method offers superiority on two important aspects. First, the convolutional neural network acts as an optimized TF kernel which, unlike any existing kernel, is trained to ensure crossterm-free TFR. Second, the deep network architecture in the DCNN provides high flexibility to optimize the TF kernel with a low complexity.

\subsection{DCNN Architectures}

The proposed DCNN structure is depicted in Fig.\  \ref{CNN_structure2}. The input to the DCNN is the 2-D WVD image $\mathbf{X}$. The corresponding crossterm-free TFR $\mathbf{Y}$ is used as the training label, which is constructed from the IF law of the signal components scaled by their respective power.

Given an $N$-layer DCNN, there are two types of layers. \begin{itemize}
    \item Conv+ReLU: For the first layer, $C$ filters of size $D \times D \times 1$  are used to generate $C$ feature maps. Rectified linear unit (ReLU, $\max (0, \cdot)$) is followed to introduce the  nonlinearity. It also pushes negative output to actual zeros, thus helping impose the TF sparsity. For layer $n=2,\cdots, N-1$, $C$ filters of size $D \times D \times C$ are utilized.
    \item Conv: For the last layer, a single filter of size $D \times D \times C$ is utilized to reconstruct the output TFR. No active function is included in the last layer so as not to limit the range of the output values.

\end{itemize}
 Denote ${\mathbf W}^{[n]}_c$ and $b^{[n]}_c$ as the weight coefficient matrix and the bias of the $c$th channel at the $n$th layer. Then, the sequential updating procedures at the different layers are given by:
\begin{equation}
\begin{split}
    \mathcal{L}^{[0]}(\mathbf{X}) &= \mathbf{X},\\
    \mathcal{L}^{[n]}_c(\mathbf{X}) &= \text{ReLU}\left(\mathbf{W}_c^{[n]} \cdot \mathcal{L}^{[n-1]}(\mathbf{X})+b^{[n]}_c\right), \\
    & \quad \  n=1,\cdots, N-1, \ c=1,\cdots, C, \\
    \mathcal{L}^{[N]}(\mathbf{X}) &= \mathbf{W}^{[N]} \cdot \mathcal{L}^{[N-1]}(\mathbf{X})+b^{[N]} = \hat{\mathbf{Y}},
\end{split}
\end{equation}
where $\mathcal{L}^{[n]}({\mathbf x})$ denotes the stacked collection of $\mathcal{L}_1^{[n]}({\mathbf x}), \cdots, \mathcal{L}_C^{[n]}({\mathbf x})$.

We fix the convolutional stride as $1$, and zero-padding is employed to keep the size of feature maps unchanged after each convolution. The receptive field size is  $(DN-N+1) \times (DN-N+1)$. Increasing the receptive field size can exploit context information in a larger TFR region.

\subsection{Neural Network Training}
As examples, we consider two-component FM signals, expressed as:
\begin{equation}
	x(t) = e^{\jmath \phi_1(t)} + e^{\jmath \phi_2(t)}
\end{equation}
for $t=0, 1,\cdots,T-1$. We assume $T=128$, and the resulting size of each input TFR image is $128 \times 128$. Two types of signals are considered. The first one consists of two NLFM signal components, whereas the second one consists of a linear FM (LFM) component and a sinusoidal FM (SFM) component. 1,500 samples are randomly generated for each class with different parameters, such as the respective initial frequencies, frequency slope and spacing. $80\%$ of the samples are utilized for training and the remaining $20\%$ are utilized for validation.

We adopt the mean square error between the estimated TFR $\hat{\mathbf{Y}}$ and $\mathbf{Y}$ as the loss function, described as
\begin{equation}
    \text{Loss}^{\{i\}} = \frac{1}{2M} \sum_{m=1}^{M} \| \hat{\mathbf{Y}}_m^{\{i\}}-  \mathbf{Y}_m^{\{i\}} \|_F^2,
\end{equation}
where $M$ is the batch size of the $i$-th batch.

In this letter, we empirically set $N=12$, $C = 40$, $D=5$, and $M = 32$ that well balance the complexity and the performance. The optimizer implements the Adam algorithm \cite{Adam}, with all its hyper-parameters set to their default values.

To verify the noise robustness of our proposed method, we consider 4 noise levels, i.e., noise-free (``inf"), $10$ dB, $5$ dB, and $0$ dB. The same parameters are shared to generate the training dataset at different noise levels.

	\begin{figure*}[!t]
		\centering	
		\footnotesize
		\subfigure{\label{fig:nocross_nlfm_label}
			\includegraphics[scale = 0.26]{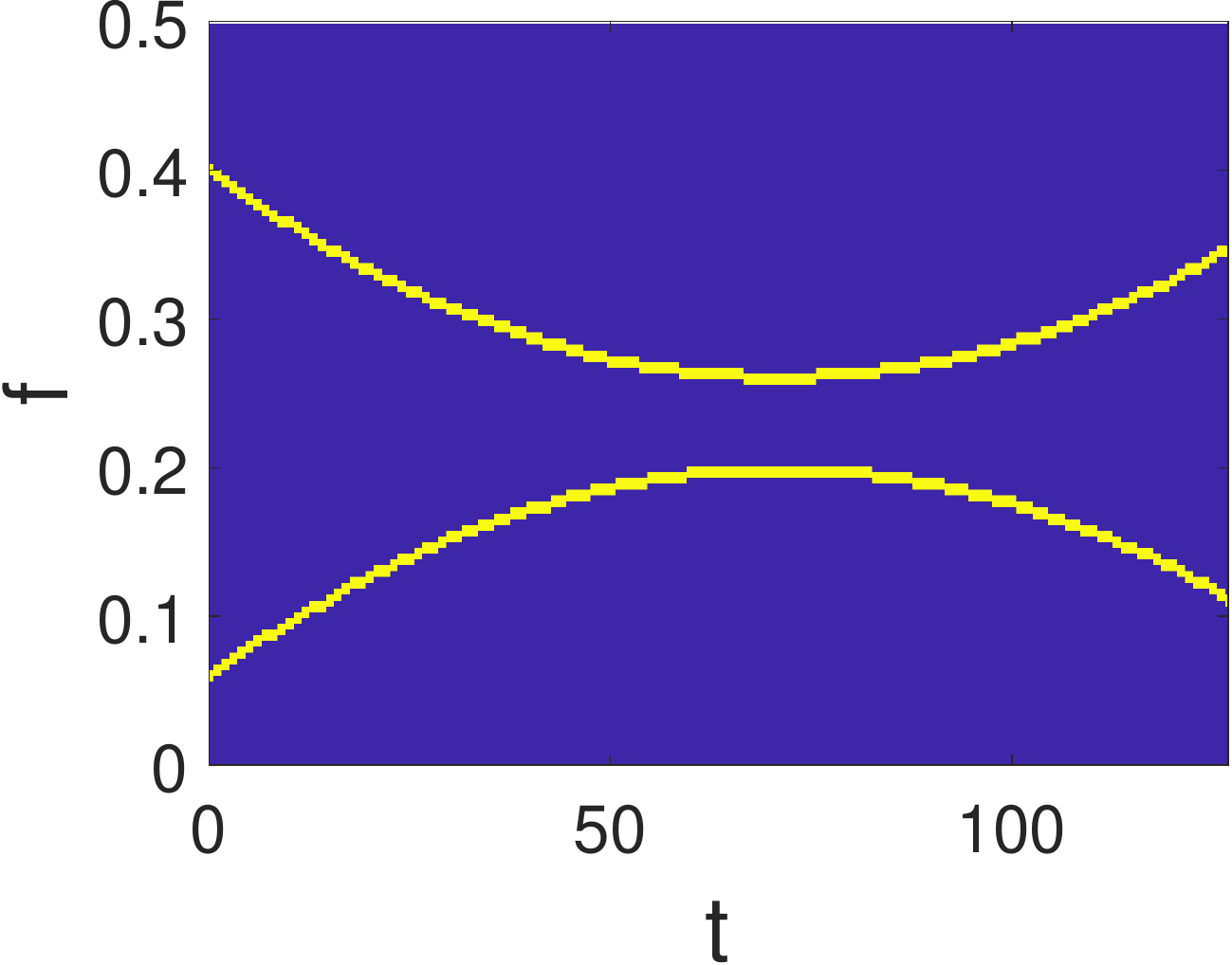}}
		\subfigure{\label{fig:nocross_nlfm_WVD}
			\includegraphics[scale = 0.26]{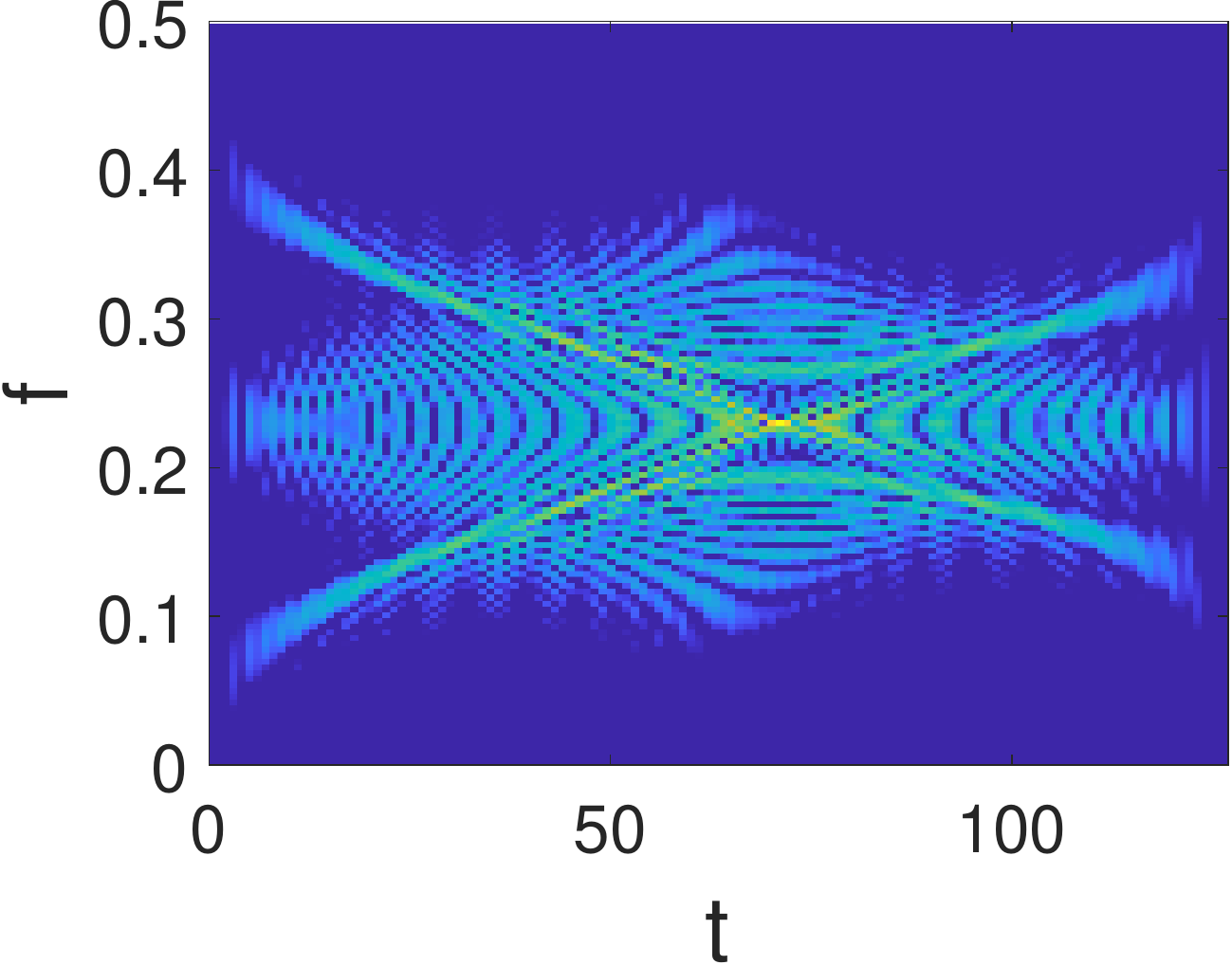}}
		\subfigure{\label{fig:nocross_nlfm_AOK_OMP}
			\includegraphics[scale = 0.26]{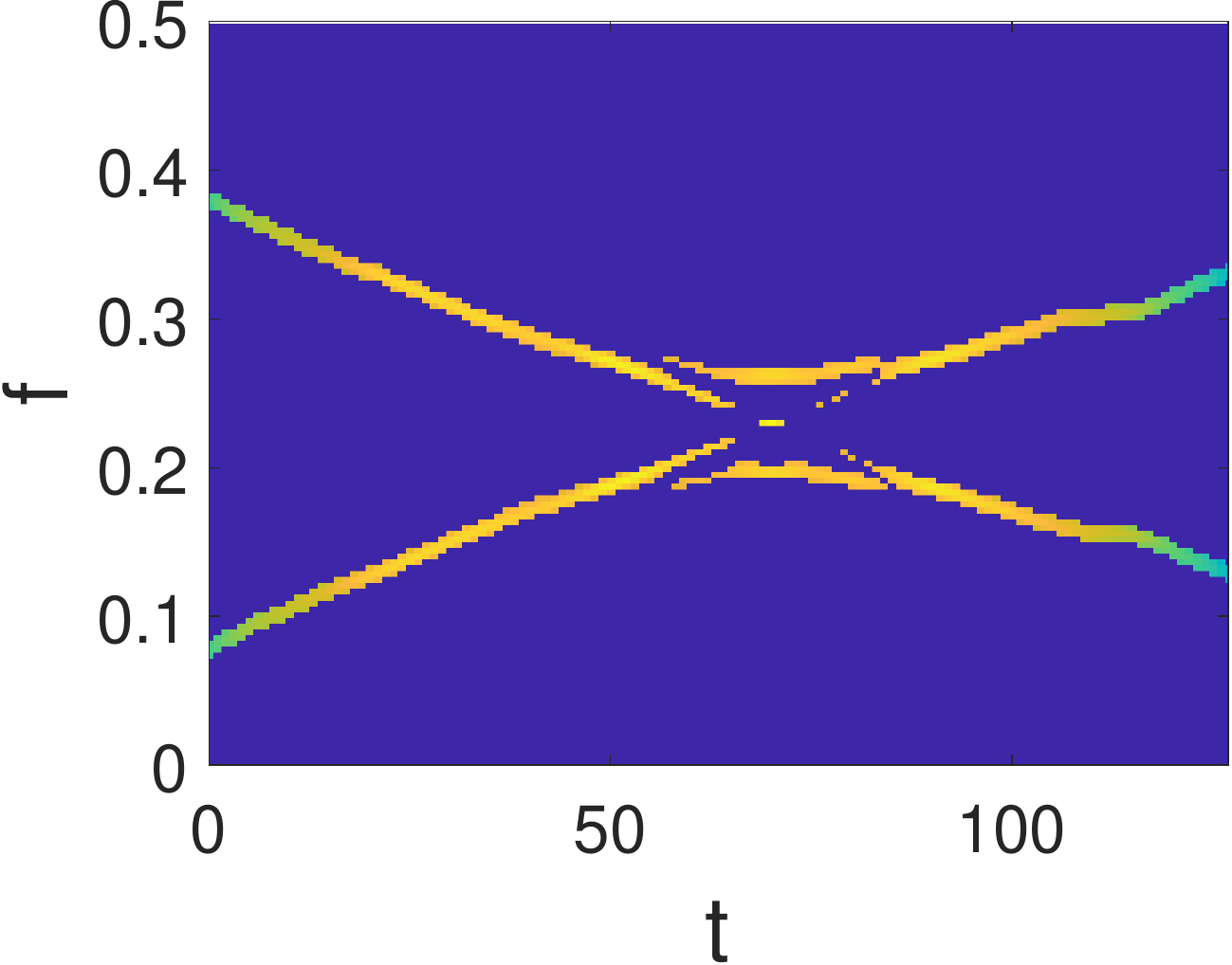}}
		\subfigure{\label{fig:nocross_nlfm_l1pox}
			\includegraphics[scale = 0.26]{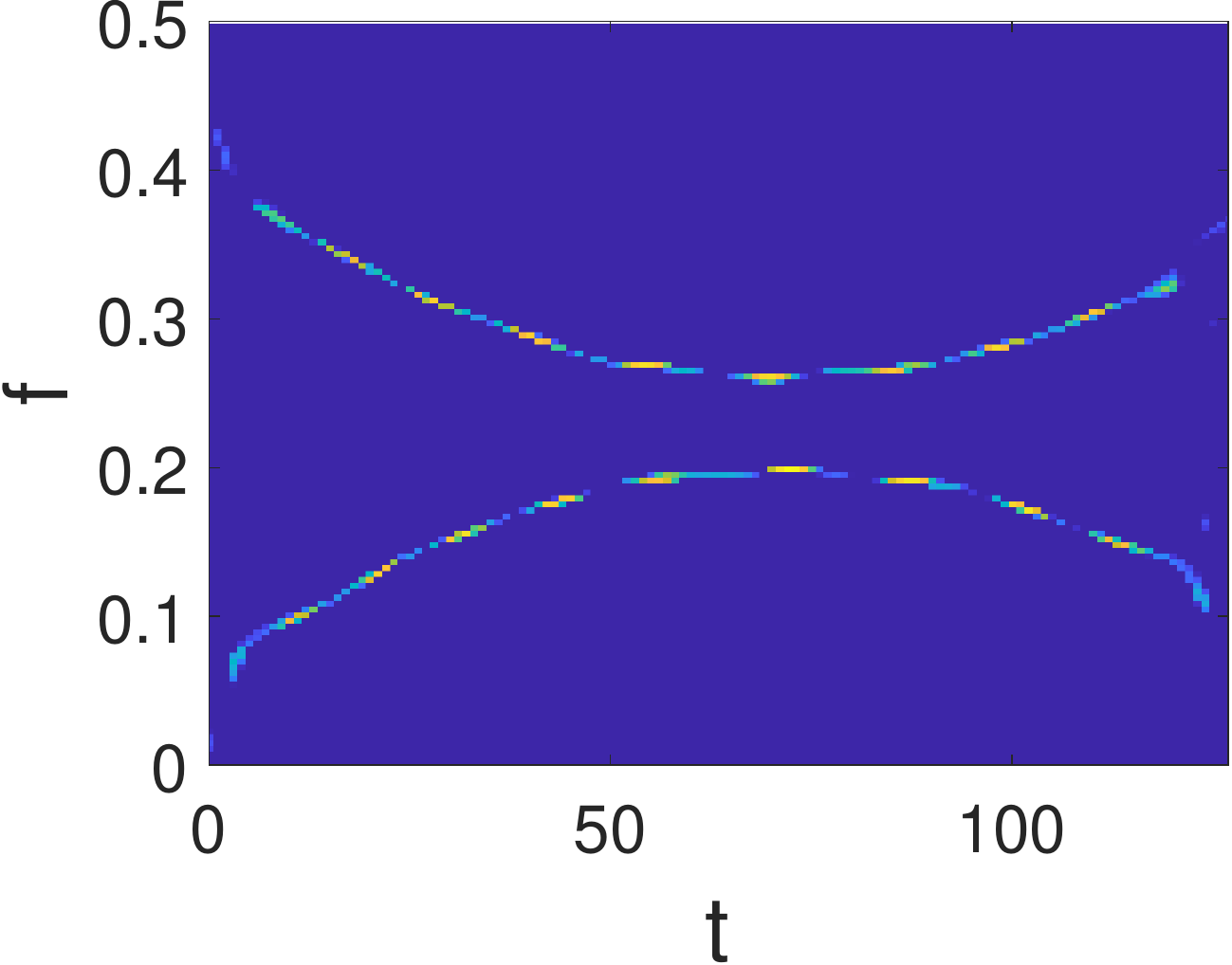}}
		\subfigure{\label{fig:nocross_nlfm_DNN}
			\includegraphics[scale = 0.26]{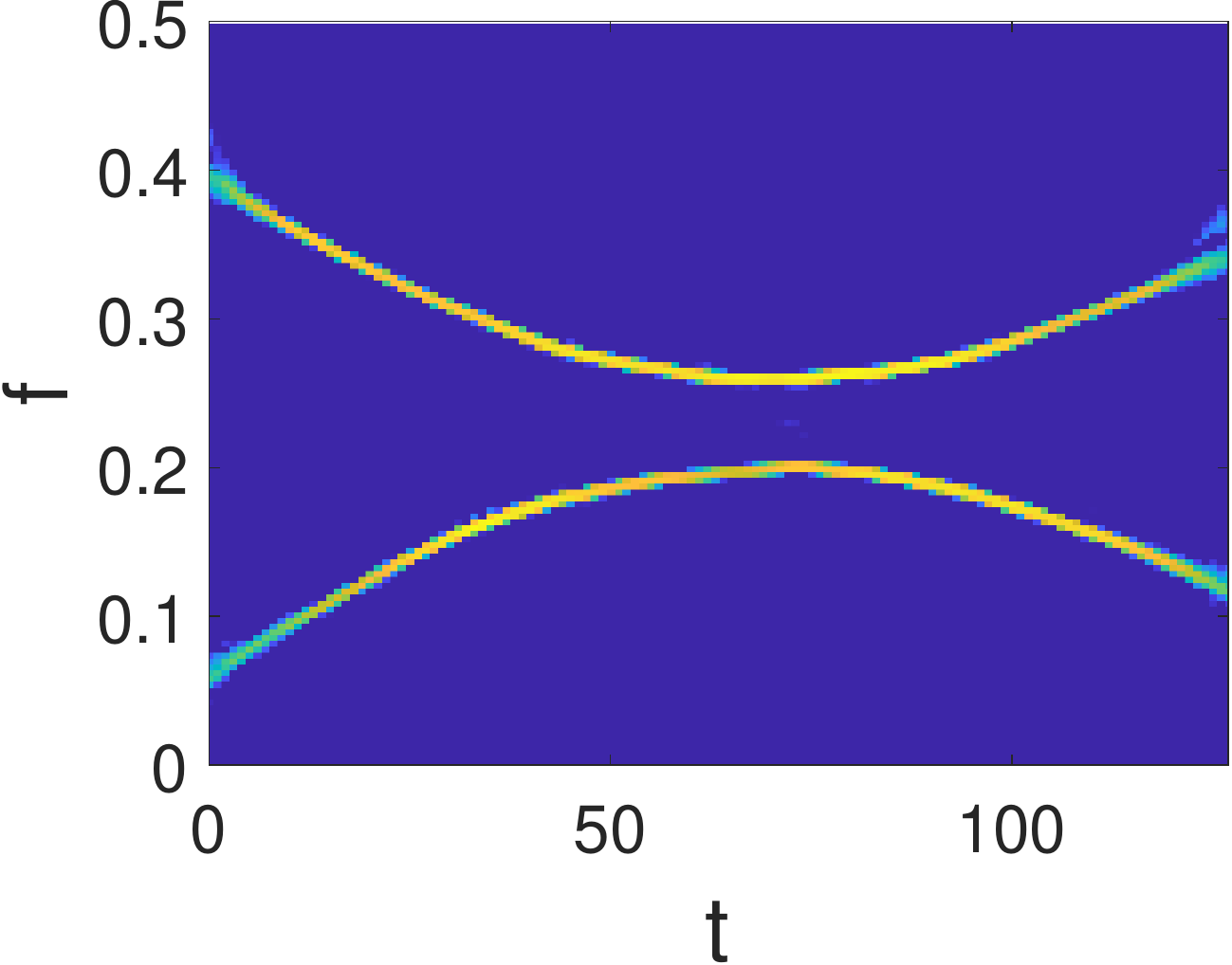}}
			\\
		\hspace{2em} (a) \hspace{11.2em}  (b) \hspace{11.2em}  (c)\hspace{11.2em} (d)\hspace{11.2em} (e)\hspace{11.2em}\\	\vspace{-0.5em}
		
		\subfigure{\label{fig:cross_nlfm_label}
			\includegraphics[scale = 0.26]{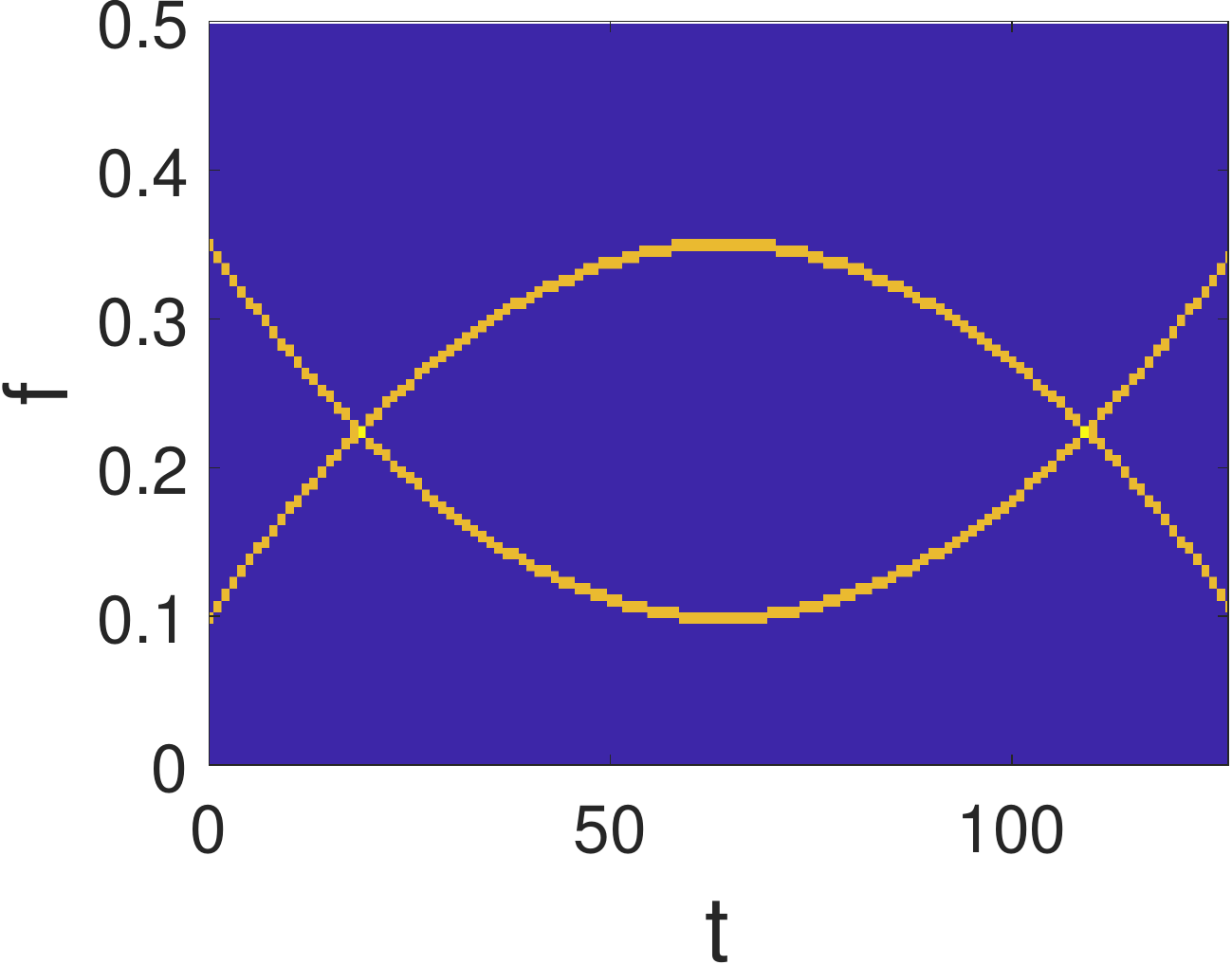}}
		\subfigure{\label{fig:cross_nlfm_WVD}
			\includegraphics[scale = 0.26]{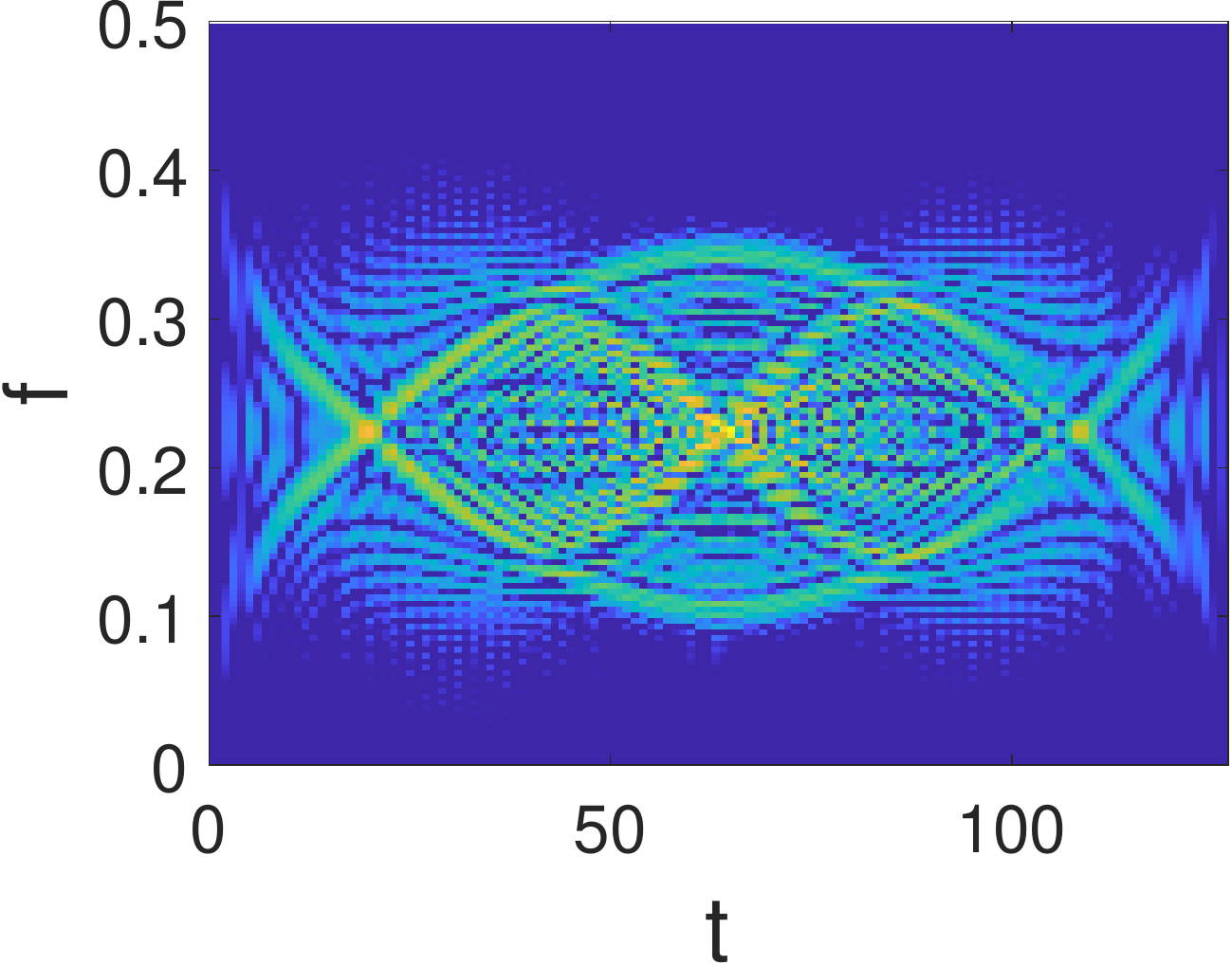}}
		\subfigure{\label{fig:cross_nlfm_AOK_OMP}
			\includegraphics[scale = 0.26]{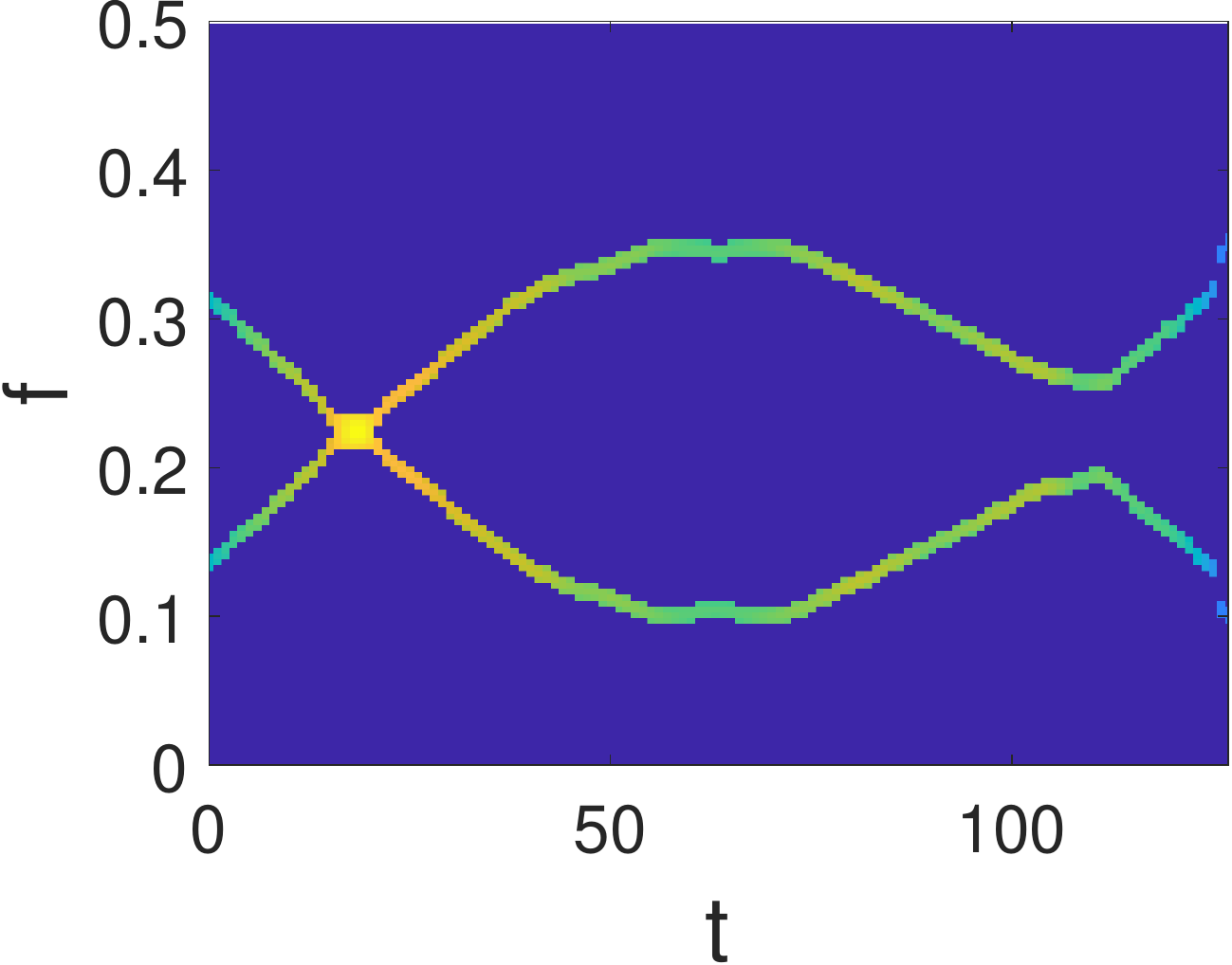}}
		\subfigure{\label{fig:cross_nlfm_l1}
			\includegraphics[scale = 0.26]{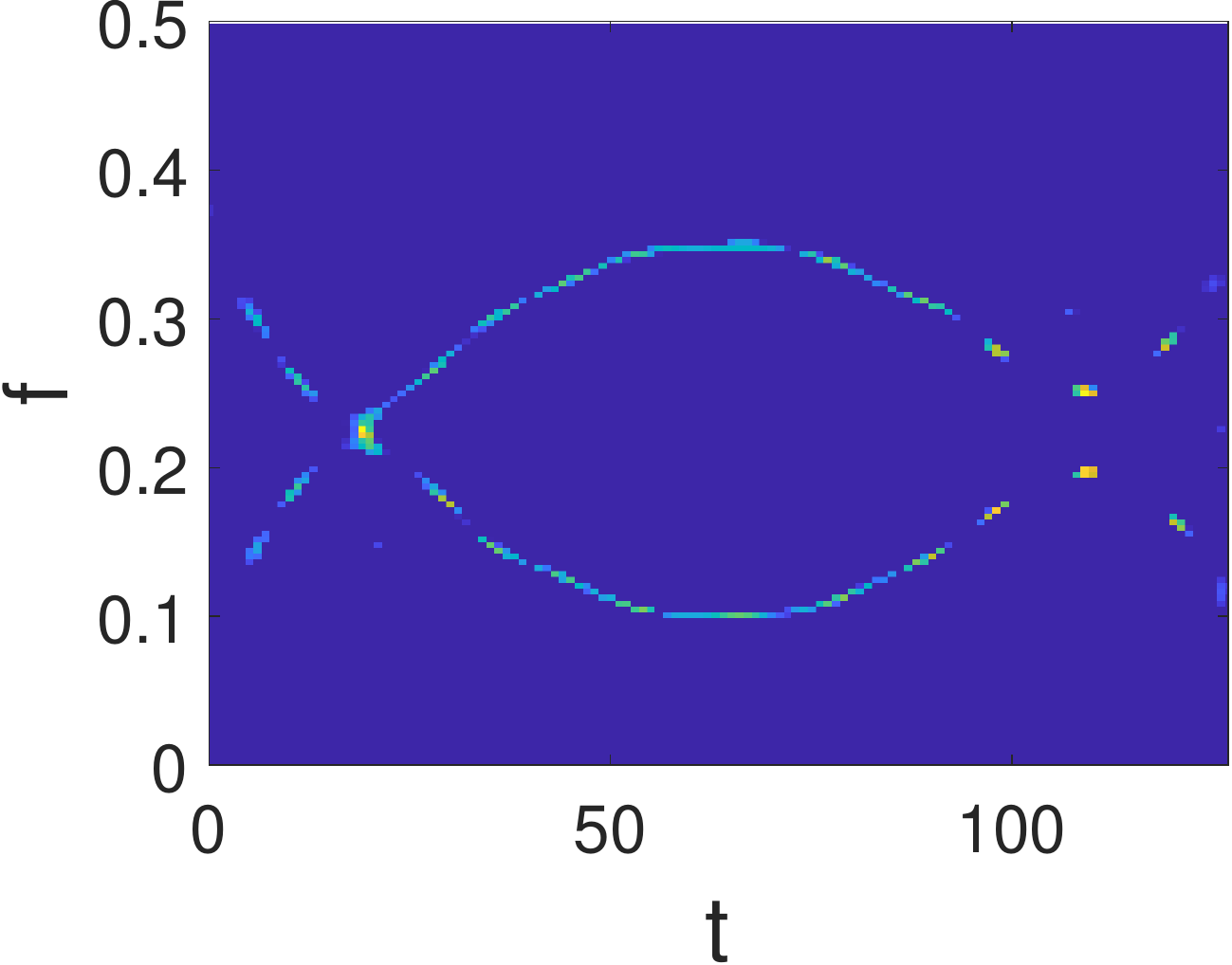}}
		\subfigure{\label{fig:cross_nlfm_DNN}
			\includegraphics[scale = 0.26]{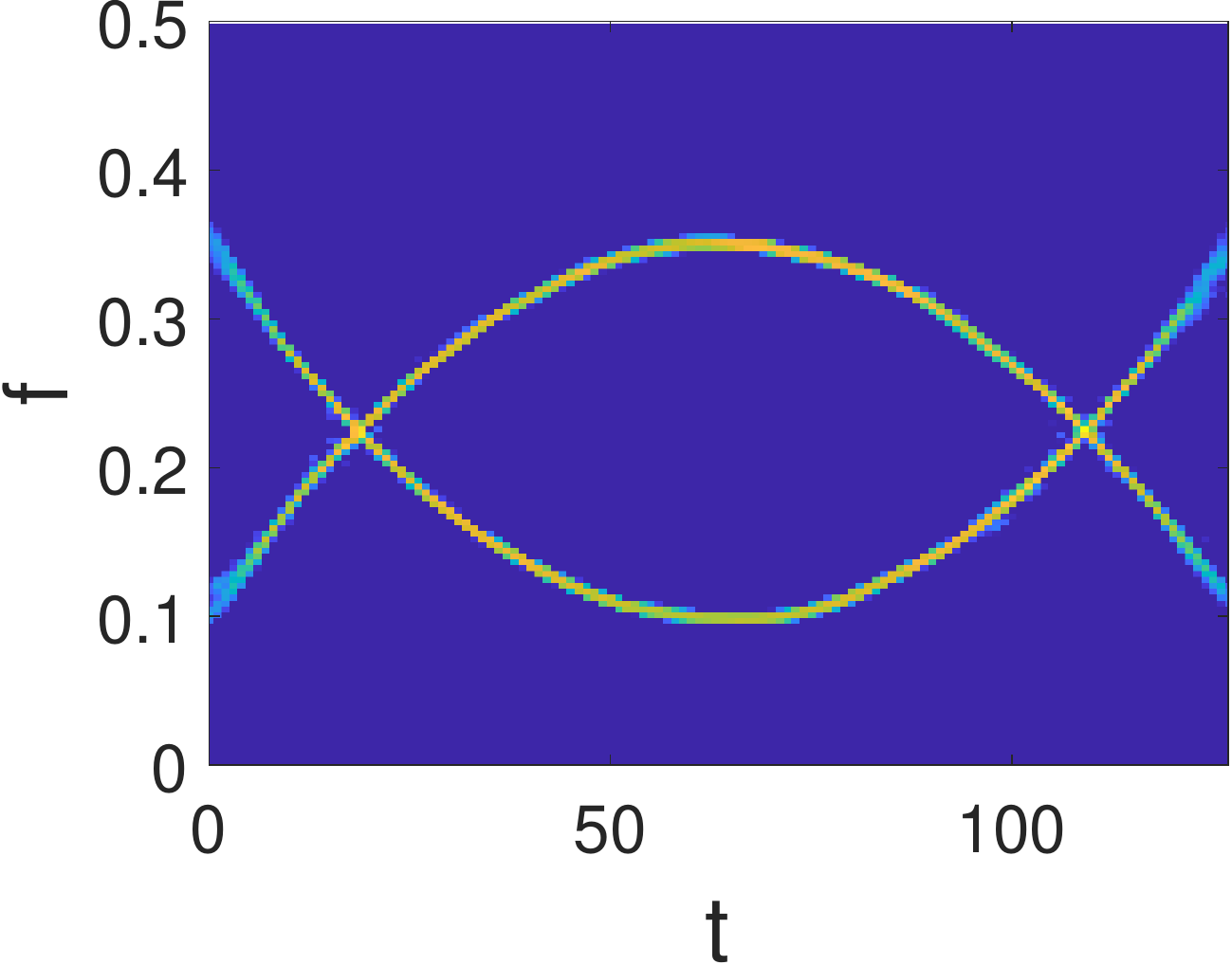}} 	\\
		\vspace{-0.5em}

	\hspace{2em} (f) \hspace{11.2em}  (g) \hspace{11.2em}  (h)\hspace{11.2em} (i)\hspace{11.2em} (j)\hspace{11.2em}\\	
		\subfigure{\label{fig:sine_label}
			\includegraphics[scale = 0.26]{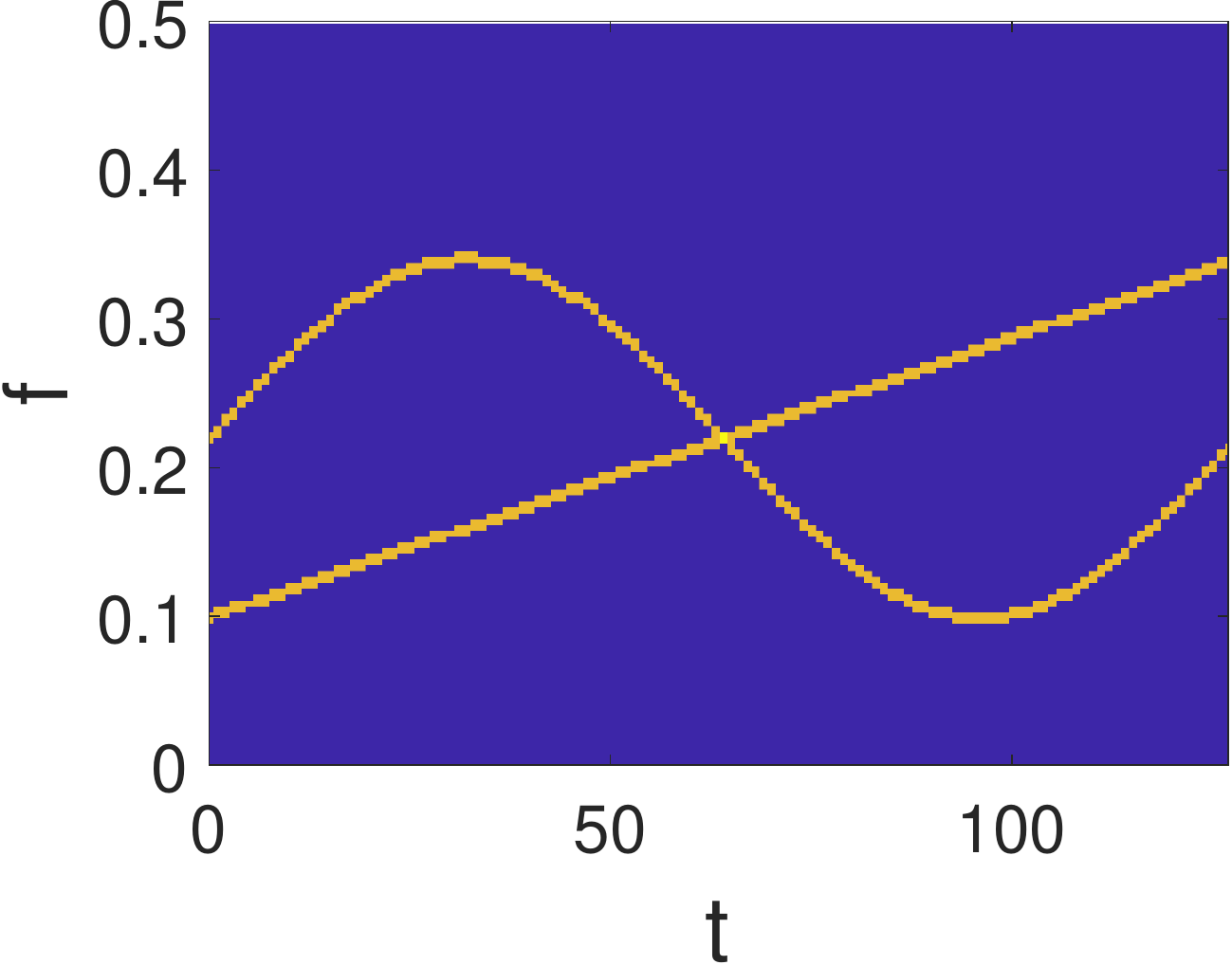}}
		\subfigure{\label{fig:sine_WVD}
			\includegraphics[scale = 0.26]{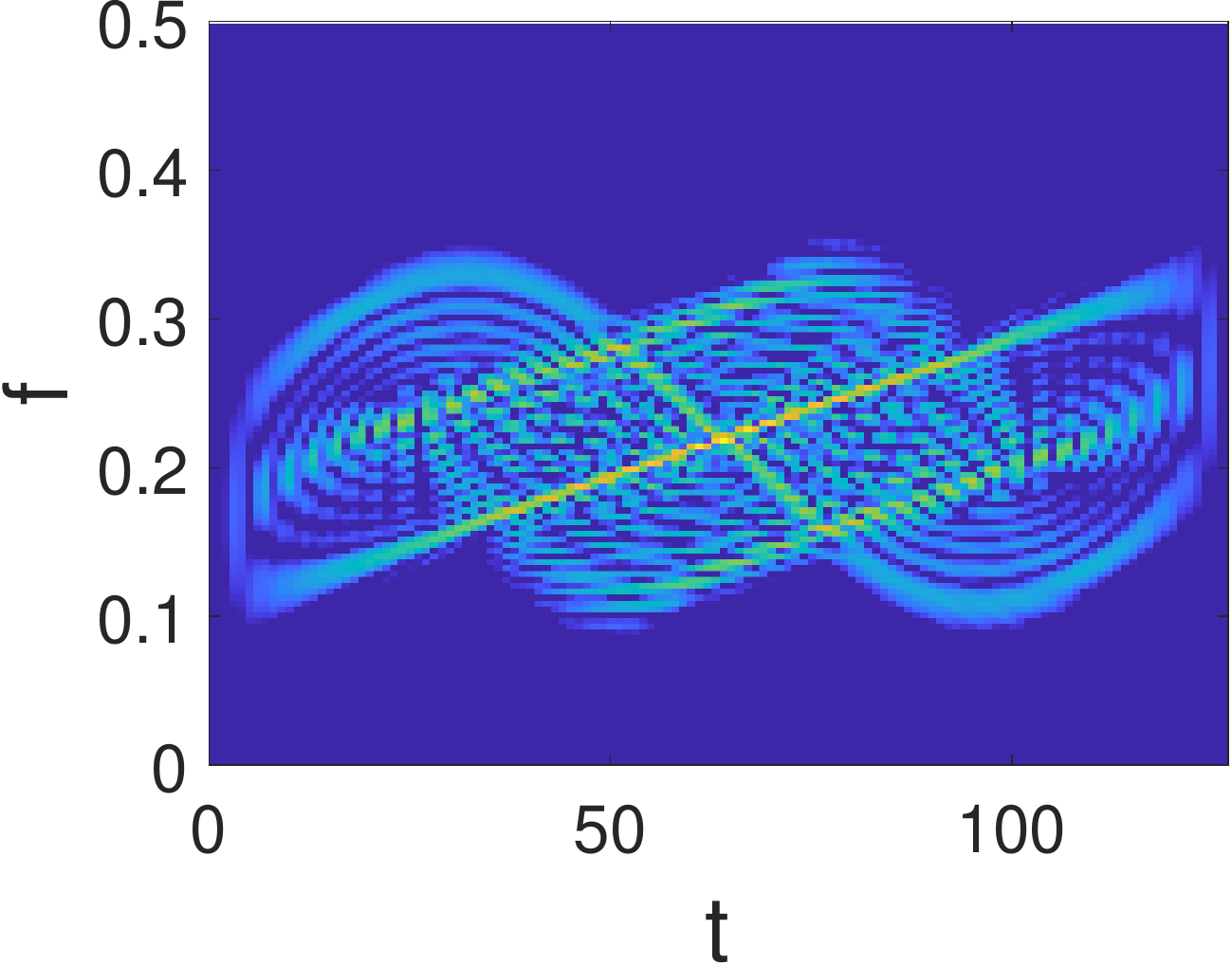}}
		\subfigure{\label{fig:sine_AOK_OMP}
			\includegraphics[scale = 0.26]{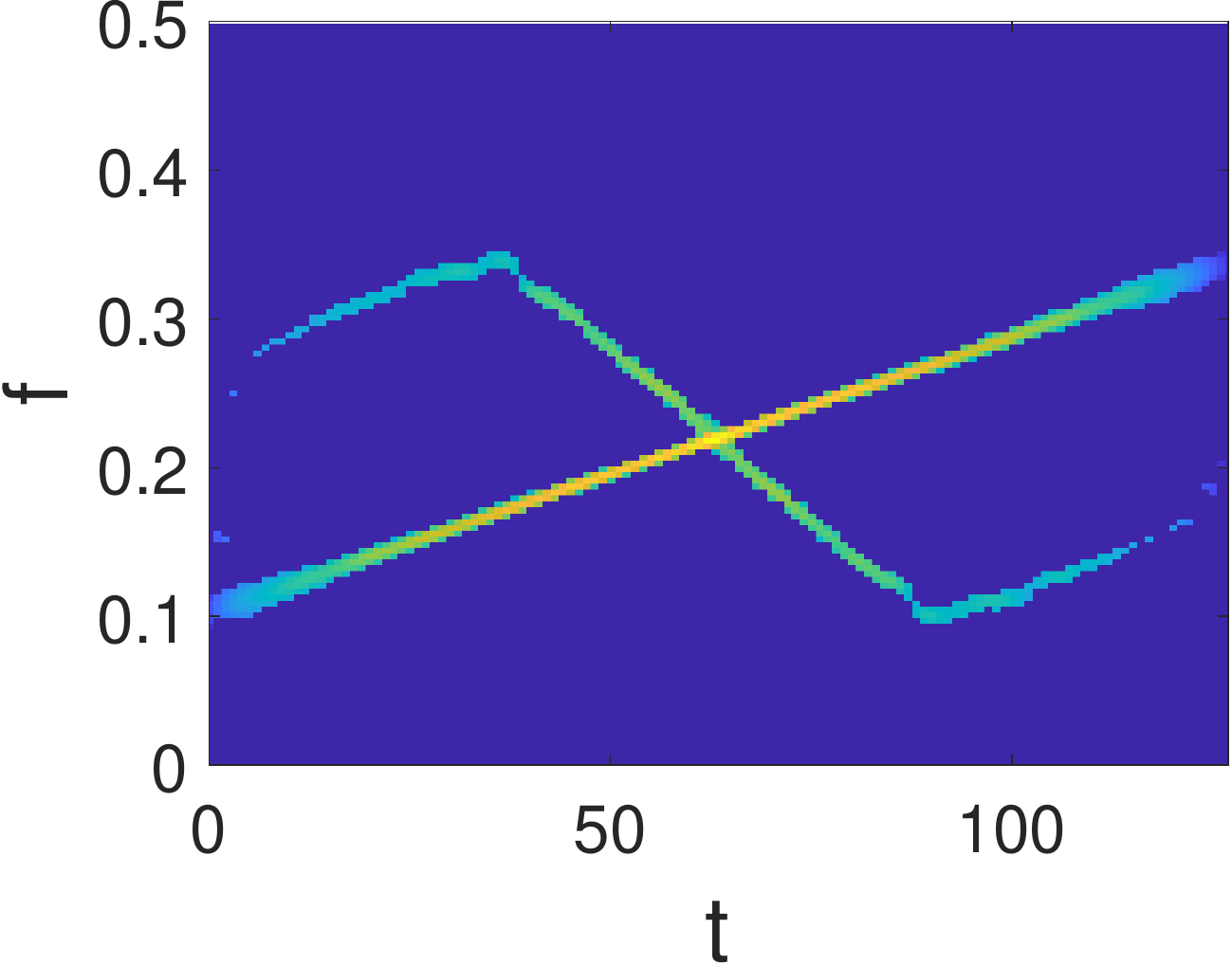}}
		\subfigure{\label{fig:sine_l1}
			\includegraphics[scale = 0.26]{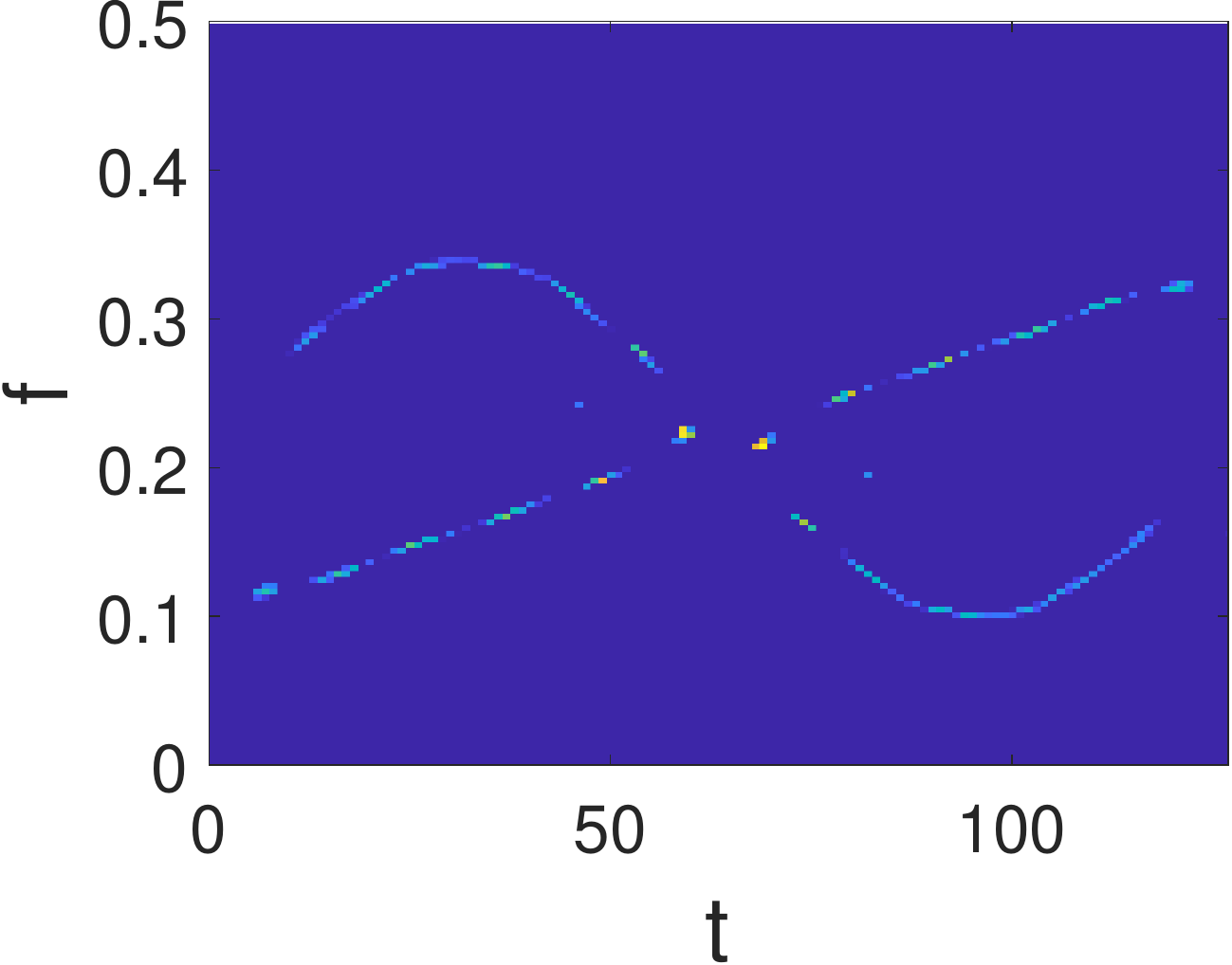}}
		\subfigure{\label{fig:sine_DNN}
			\includegraphics[scale = 0.26]{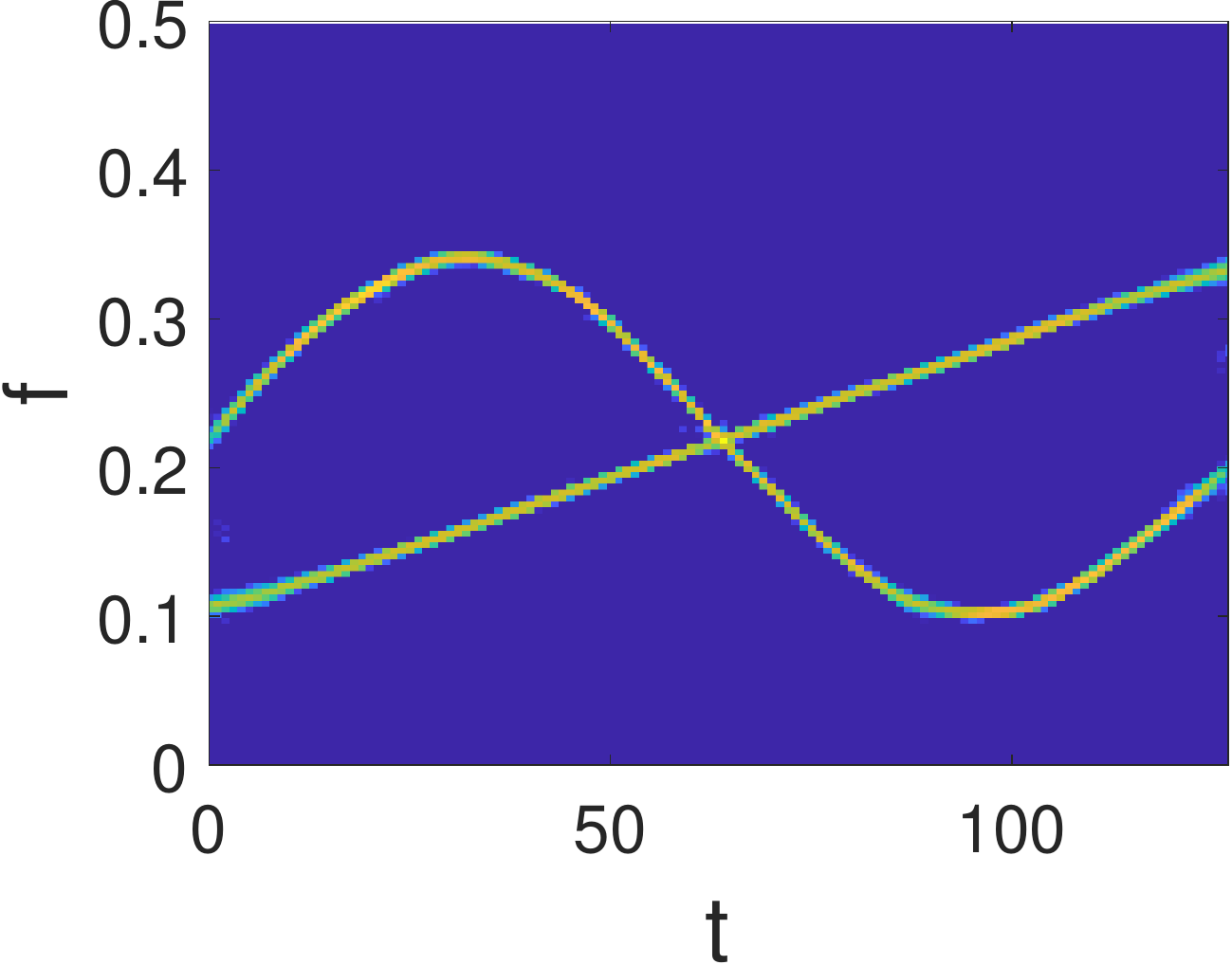}}\\
		
		\hspace{2em} (k) \hspace{11.2em}  (l) \hspace{11.2em}  (m)\hspace{11.2em} (n)\hspace{11.2em} (o)\hspace{11.2em}\\	
		\vspace{-0.5em}
		
		\caption{Reconstructed TFR results for the three cases at input $\text{SNR}=5\ \text{dB}$. From top to bottom: Case 1 to Case 3. From left to right: Ideal TFR, WVD, AOK+OMP, $\ell_1$-prox \cite{Flandrin2015}, and the proposed TFR. For all diagrams, amplitudes are coded logarithmically with a dynamic range of 15 dB.}
		\vspace{-0.5em}
		\label{fig:Performance_comparison1}
	\end{figure*}

	\begin{table*}[t]
	\centering
	\caption{NMSE ({\rm dB}) comparison among WVD, AOK+OMP \cite{Yimin_reduced}, $\ell_1$-prox \cite{Flandrin2015} and the proposed DCNN-based method for different SNR values}
	\vspace{-1.5em}
	\label{table_2}
		\begin{center}
			\begin{tabular}{c   c c c c   c  c c c c  c  c c c c}
				\toprule
				\multirow{2}*{SNR} &\multicolumn{4}{c}{Case 1: Two closely located NLFMs} & & \multicolumn{4}{c}{Case 2: Two overlapped NLFMs}  & & \multicolumn{4}{c}{Case 3: Overlapped LFM and SFM}\\
				\cmidrule{2-5} \cmidrule{7-10} \cmidrule{12-15}
               (dB) & WVD & \!\!\!AOK+OMP\!\!\! & $\ell_1$-prox & Proposed  && WVD & \!\!\!AOK+OMP\!\!\! & $\ell_1$-prox & Proposed && WVD & \!\!\!AOK+OMP\!\!\! & $\ell_1$-prox & Proposed\\
				\midrule
				Inf   &   $0.06$   & $-1.38$  & $-0.50$ &  $-9.65$      & &  $-0.45$ & $-1.68$ & $-1.03$  & $-7.27$  & &  $2.57$ & $-0.30$ & $-0.76$  &$-8.36$ \\
				$10$    &   $0.35$   & $-1.36$   & $-0.41$ &  $-6.29$      & & $ -0.38$  & $\ \ \ 1.80$ & $-0.90$  &$-3.94$  & &  $2.92$  & $\ \ \ 2.59$ & $-0.57$  &$-6.40$ \\
				$5$     &   $0.93$   & $\ \ \ 1.90$   & $-0.29$ &  $-4.69$      & &  $-0.21$  & $\ \ \ 1.70$ & $-0.64$  & $-3.86$  & &  $3.58$  & $\ \ \ 2.76$ & $-0.38$  & $-5.09$ \\
				$0$     &   $2.56$   & $-1.14$   & $-0.18$ &  $-2.78$     & &  $\ \ \ 0.39$  & $-1.03$ & $-0.38$  &$-2.23$  & &  $5.14$  & $\ \ \ \ 1.04$ & $-0.23$  & $-2.23$\\
				\bottomrule
			\end{tabular}
		\end{center}
	\vspace{-2.0em}
\end{table*}

\section{Simulation Results}

We compare our proposed DCNN-based method with commonly used state-of-the art algorithms, the WVD, AOK+ OMP \cite{Yimin_reduced}, and $\ell_1$-prox \cite{Flandrin2015}. The $\ell_1$-prox algorithm \cite{Flandrin2015} takes $13 \times 13$ sampling rectangular area near the AF origin.

\subsection{Case Studies}

We consider three different cases with NLFM, LFM and SFM components. The three cases involve the complex structure, e.g., closely-located and crossing components. Visual comparison of different TFR reconstruction methods obtained at input signal-to-noise ratio (SNR) of $5\ \text{dB}$ is depicted in Fig.\ \ref{fig:Performance_comparison1}. In Fig.\ \ref{fig:Performance_comparison1}, row 1 to row 3 are respectively associated with case 1 to case 3.
\subsubsection{Case 1: Two closely located NLFMs}
We first consider the following two-component NLFM signal with their phase laws given as:
\begin{equation}\label{nocross_NLFM}
\begin{split}
	\phi_1(t) &= 2\pi \left(0.06t + 0.25t^2/T - 0.15t^3/T^2\right),\\
	\phi_2(t) &= 2\pi \left(0.40t - 0.25t^2/T + 0.15t^3/T^2\right).
\end{split}
\end{equation}
 The model TFR and the WVD are representatively presented in Fig.\ \ref{fig:nocross_nlfm_label} and Fig.\ \ref{fig:nocross_nlfm_WVD}. Fig.\ \ref{fig:nocross_nlfm_AOK_OMP} shows the TFR reconstruction by AOK+OMP. We notice that AOK+OMP cannot separate the two signal components when they get close ($58\leq t \leq 80$). As shown in Fig.\ \ref{fig:nocross_nlfm_l1pox}, $l_1$-prox is able to separate the two components. However, the IFs are not continuous especially for their middle portions. Because $\ell_1$-prox tends to impose the global sparsity in the TF domain instead of the local sparsity. Fig.\ \ref{fig:nocross_nlfm_DNN} shows the the reconstructed TFR via the proposed DCNN-based method, which is very close to the model TFR.

\subsubsection{Case 2: Two overlapped NLFMs}	
The instantaneous phase laws of the two components are respectively given as:
	\begin{equation}\label{cross_NLFM}
     \begin{split}
     \phi_1(t) &= 2\pi\left(0.35t - 0.50t^2/T + 1/3 t^3/T^2 \right),\\
     \phi_2(t) &= 2\pi\left(0.10t + 0.50t^2/T - 1/3 t^3/T^2\right).
     \end{split}
   \end{equation}	
 Compared to case 1, the signal consisting of two component NLFM with crossing points is much more challenging to handle. For AOK+OMP, we notice that the crossterms are greatly eliminated. However, AOK+OMP is not able to detect the spectral overlapped portion when $t=110$, as depicted in Fig.\ \ref{fig:cross_nlfm_AOK_OMP}. For $\ell_1$-prox, compared with AOK+OMP, it has less IF distortions especially in the middle part, as shown in Fig.\ \ref{fig:cross_nlfm_l1}. However, $\ell_1$-prox still fails to reconstruct IFs in the second overlapped portion. Because in such as a case, the autoterms and the crossterms are difficult to be separated in the AF domain. Fig.\ \ref{fig:cross_nlfm_DNN} show the TFR obtained by our proposed method. We notice that the proposed method is the only one to accurately restore the information at the intersections.

 \subsubsection{Case 3: Overlapped SFM and LFM}
 In this case, the instantaneous phase laws of the two components are:
	\begin{equation} \label{SFM and LFM}
	\begin{split}
	\phi_{1}(t) &=  2\pi \left((0.06T/\pi) \cos \left(2\pi t/T + \pi\right)+0.22t\right),\\
	\phi_{2}(t) & = 2\pi \left(0.10t + 0.12t^2/T\right).
	\end{split}
	\end{equation}
Fig.\ \ref{fig:sine_AOK_OMP} depicts the TFR reconstructed by AOK+OMP, where the LFM component is well reconstructed, but the SFM component is heavily distorted since it is difficult for the AOK to handle highly nonlinear IF signatures. Fig.\ \ref{fig:sine_l1} shows the reconstructed result via $\ell_1$-prox. Compared with AOK+OMP, $\ell_1$-prox is more capable to handle the high IF nonlinearity. However, similar to Fig.\ \ref{fig:cross_nlfm_l1}, the IFs get deformed at the overlapped portion. Fig.\ \ref{fig:sine_DNN} presents our proposed TFR, which provides a desirable TF resolution with all IF information restored correctly.

In all the above three cases, the proposed method consistently provides near-ideal TFRs irrespective of the signal types, and even when the input SNR is relatively low ($\text{SNR}=5\ \text{dB}$). In particular, the proposed method is rarely affected by the finite sampling effect and maintains the signal energy well along the signal IFs.

\vspace{-0.5em}
\subsection{Robustness Analysis}
To quantitatively compare the performance of the proposed DCNN-based method with the WVD, AOK+OMP \cite{Yimin_reduced}, and $\ell_1$-prox \cite{Flandrin2015}, we use the averaged normalized mean square error (NMSE) as the  performance assessment criterion:
\begin{equation}
\vspace{-0.25em}
 \mathrm{NMSE}=\frac{1}{K} \sum_{k=1}^K 10 \log _{10}\left(\| \mathbf{Y}-\hat{\mathbf{Y}_k}\|_{2}^{2} / {\|\mathbf{Y}\|_{2}^{2}}\right).
\end{equation}	
We consider noisy signal measurements with different levels of input SNR. $K=50$ independent trials are performed to average the NMSE for each scenario. Table \ref{table_2} compares the NMSE results in dB between the proposed method and the three existing methods. Generally, the TFR reconstruction performance is improved as the SNR increases. However, for AOK+OMP, the NMSE does not necessarily decrease with the input SNR since the AOK fails to reconstruct a high-fidelity TFR in all three challenging cases even when no noise is present. We notice in Table \ref{table_2} that our proposed method consistently provides the lowest NMSE in all listed scenarios. In particular, the NMSE of the proposed method obtained at $\text{SNR} = 0\ \text{dB}$ is lower than the NMSE of the other methods obtained for noise-free signals. The effectiveness and robustness of the proposed method is thus evidently verified.

\vspace{-0.5em}
\section{Conclusion}
In this letter, we proposed a DCNN-based method to obtain crossterm-free TFRs. In the proposed method, a deep neural network is trained to provide effective discriminative learning capability and completely mitigate undesired crossterms while preserving signal autoterms.
The performance of the proposed method is not restricted by the geometrical shapes in any domain as in the conventional TF kernel design.
The effectiveness and robustness of the proposed method are examined via numerical simulations in terms of the reconstructed TFRs and the NMSE performance. The proposed method provides significant performance improvement compared to existing TFR reconstruction algorithms, including the classical WVD as well as state-of-the-art AOK+OMP and $\ell_1$-prox. The proposed method works robustly even when the signals consist of closely located or spectrally overlapping components in noisy environments.

\end{document}